\documentclass{elsart}

\usepackage{graphicx}

\usepackage{amssymb,amsmath,bbm}

\begin{document}

\begin{frontmatter}

\title{Naimark-Sacker Bifurcations in Linearly Coupled Quadratic Maps}

\author[SC,PR]{Paulo C.~Rech\corauthref{cor}},
\corauth[cor]{Corresponding author}
\ead{rech@fisica.ufpr.br}
\author[PR]{Marcus W.~Beims},
\author[PR,RS,BRD]{Jason A.~C.~Gallas}
\address[SC]{Departamento de F\'\i sica, 
             Universidade do Estado de Santa Catarina,\\
             89223-100 Joinville, Brazil.}

\address[PR]{Departamento de F\'\i sica, Universidade Federal do Paran\'a,\\
             81531-990  Curitiba, Brazil.}

\address[RS]{Instituto de F\'\i sica, 
             Universidade Federal do Rio Grande do Sul,\\
              91501-970 Porto Alegre, Brazil.}

\address[BRD]{Institut f\"ur Computer Anwendungen, Universit\"at Stuttgart,\\
             Pfaffenwaldring 27, D-70569 Stuttgart, Germany.}
\begin{abstract}
We report exact analytical expressions locating the
$0\to1$, $1\to2$ and $2\to4$ bifurcation curves
for a prototypical system of two linearly coupled quadratic maps.
Of interest is the precise location of the parameter
sets where Naimark-Sacker bifurcations occur, 
starting from a non-diagonal period-$2$ orbit. 
This result is the key to understand the onset of synchronization
in networks of quadratic maps.
\end{abstract}


\end{frontmatter}


An essential clue to understand how dynamical systems synchronize and
the origin of their riddled basins of attraction is provided by the
mechanisms underlying the transverse desestabilization of low periodic
orbits embedded in chaotic 
attractors\cite{sync,bocca,livrokurths,maistrenko98}.
So far, these mechanisms have been investigated only via direct numerical 
simulation of the dynamics. 
The purpose of this paper is to report exact analytical results for
a prototypic model, two linearly coupled quadratic maps, defined by
the following pair of equations
\begin{equation}
   x_{t+1} = a - x_t^2 + b\;(x_t - y_t),  \qquad\quad
   y_{t+1} = a - y_t^2 + b\;(y_t - x_t).\label{model}
\end{equation}
Here $x_{t}$, $y_{t}$ represent dynamical variables, 
$a$ is the local nonlinearity, $b$ the coupling, and $t=0,1,2,\dots$ is 
the discrete time. 
Systems of coupled nonlinear oscillators are nowadays used in practical 
appli\-cations, e.g., in the investigation of coupled p-n 
junctions~\cite{bj} and of Josephson-junction arrays~\cite{wh}. 
They also serve as useful approximations of 
nonlinear partial differential equations describing, for example, 
the onset of fluid turbulence~\cite{agr} and 
reaction diffusion systems~\cite{wk}. 
Previous analytical results were obtained using linear stability 
analysis\cite{froy,hh,kn}, and renormalization theory\cite{kls}. 
Here we go well beyond this presenting
(i)  the algebraic varieties delimiting stability domains 
     for motions of periods $1$ and $2$, and,
(ii) the Naimark-Sacker line (also called Hopf bifurcation line)
     showing where quasiperiodicity
     stabilizes following a period-$2$ oscillation.
The motivation for considering this problem anew is our interest in using
more realistic and flexible maps as local oscillators in applications dealing
with diffusive-advective effects like, e.g.~in
ocean convection\cite{ocean1,ocean2,ocean3}.
For additional discussion of problems involving coupled systems we refer
the reader to the excellent tutorials in 
Refs.~\cite{sync,bocca,livrokurths,maistrenko98}.

As is well-known\cite{gucki,wig}, there are three ways in which a fixed point
$p$ of a discrete map $f$ may fail to be hyperbolic:
$Df(p)$ may have an eigenvalue $+1$, an eigenvalue $-1$, or a pair
of complex eigenvalues, say $\lambda$ and $\overline\lambda$, with
$\vert\lambda\vert=1$. 
The first two situations are familiar from period-doubling cascades and
appear much more frequently than the last one.
But it is precisely this last situation, characteristic of a 
Naimark-Sacker bifurcation,
that is of great interest for coupled systems and that we consider here.



\begin{figure}[tbh]
  \begin{center}
  \includegraphics*[width=14cm,angle=0]{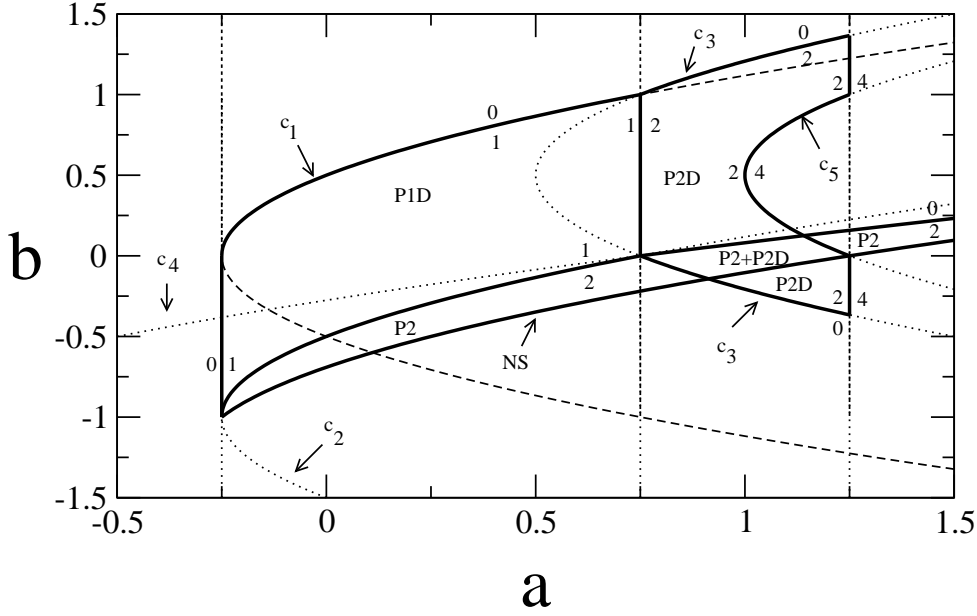}
  \end{center}
  \caption{\protect Stability domains for period-$1$ and period-$2$ 
                    orbits. $P2D$ and $P2$ refer to domains of
                    diagonal and non-diagonal orbits, respectively.
                    Numbers indicate periods.}
  \label{fig:fig1}
  \end{figure}

Equation~(\ref{model}) has four fixed points, two {\it diagonal\/}
and two {\it non-diagonal}:
\begin{equation} 
   D_1\equiv(x_1,x_1), \quad D_2\equiv(x_2,x_2), \qquad
   N_1\equiv(x_3,x_4), \quad N_2\equiv(x_4,x_3),  \label{pfixos}
\end{equation}
where $x_{1} = (-1+ \xi)/2$,
      $x_{2} = (-1- \xi)/2$, 
      $x_{3} = (2b - 1 + \eta)/2$,  
      $x_{4} =(2b - 1 - \eta)/2$, and
$\xi = \sqrt{1 + 4a}$ and $\eta=\sqrt{1+4a-4b^2}$.
The stability of these points is ruled by the equation 
$\vert J - \lambda I \vert=0$, where $\lambda$ is the eigenvalue, 
$I$ is the identity matrix and
$J$ is the Jacobian matrix of the mapping.
The eigenvalues of these four fixed points are always real,
with $D_2$, $N_1$ and $N_2$ being always unstable
since at least one of their eigenvalues is larger than $1$.
The remaining fixed point, $D_1$, is stable in the region delimited by
$-1/4<a<3/4$ and $\sqrt{1 + 4a}/2 - 1 < b < \sqrt{1 + 4a}/2$,
bounded by the curves
\[ a=-\frac{1}{4}, \quad a=\frac{3}{4},\quad
  c_1\equiv4a - 4b^2 + 1=0, \quad
  c_2\equiv4a - 4b^2 - 3 - 8b=0,  \]
shown in Fig.~\ref{fig:fig1}.

The stable fixed point bifurcates into two quite different period-$2$ 
orbits when one increases $a$, depending on the value of $b$.
When crossing the line $a=3/4$ one finds {\it diagonal\/} period-$2$ 
orbits with signature 
$(u,u) \to (v,v) \to (u,u) \to \dots$, where
$u = (1 + \sqrt{-3 + 4a})/2$ and $v = (1 - \sqrt{-3 + 4a})/2$. 
We denote such orbits by $P2D$.
Beyond the curve $c_2=0$ one finds {\it non diagonal\/} orbits with
a signature
$(s,t) \to (t,s) \to (s,t) \to \dots$ where 
$s = (2b + 1 + \sqrt{4a-4b^2 - 8b - 3})/2$ and 
$t = (2b+1-\sqrt{4a-4b^2-8b-3})/2$. These we represent by $P2$. 


The $P2D$ orbit has real eigenvalues and is stable in the domain delimited
by the curves
\[ a = \frac{3}{4}, \quad  a=\frac{5}{4},  \quad
   c_3\equiv 4a - 4b^2 - 3 + 4b=0,  \quad
   c_5\equiv 4a - 4b^2 - 5 + 4b=0.  \] 
It loses stability along $c_5=0$ or $a=5/4$,
after a $2\to4$ bifurcation. 
There is a $P4D$ diagonal orbit when crossing the line $a=5/4$, 
and a $P4$ non-diagonal orbit when crossing $c_5=0$.

The eigenvalues of the $P2$ orbit are 
\begin{equation}
     \lambda_{\pm} =  6b^2 + 10b - 4a + 4 \pm 2b\sqrt{5b^2 + 10b - 4a + 4} 
           \label{m1} 
\end{equation}  
and can be complex. 
From these expressions we obtain\cite{gucki} the following formula
for the Naimark-Sacker bifurcation line:
\begin{equation}
     a_{NS} = b^2 + \frac{5}{2}b + \frac{5}{4}.     \label{hopf}
\end{equation}

  \begin{figure}[tbh]
  \begin{center}
  \includegraphics*[width=14cm]{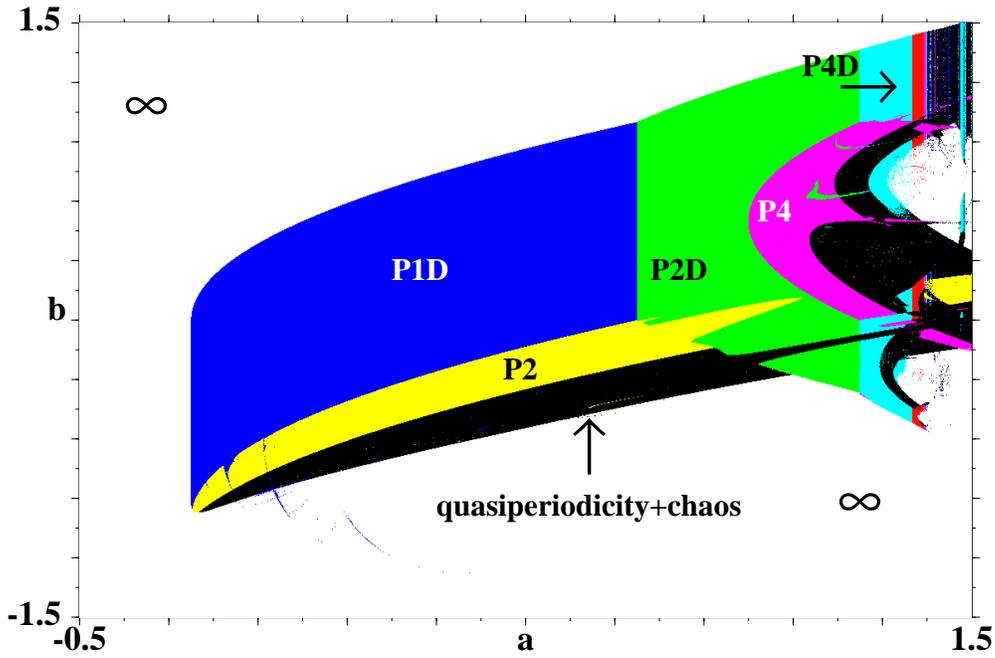}
  \end{center}
\caption{\protect Stability domains obtained by discretizing the 
         parameter interval in a mesh  of $1200\times 600$ points. 
         Different gray shadings represent stability 
                    domains of attractors with different periods.}
  \label{fig:fig2}
  \end{figure}

It is important to observe that the linear coupling of the oscillators
was fundamental in allowing the derivation of the above analytical
result. Already a quadratic coupling generates formulas which are 
intractable with present day hardware and commercially available software.

From the eigenvalues along $c_2=0$ as well as the condition 
$\lambda_1=\lambda_2=1$ we find that:
(i) for $b>0$, an unstable P2 orbit is born along $c_2=0$
    stabilizing however along $c_4=4a - 4b^2 - 3 - 12b=0$, and 
(ii) for $-1<b<0$, a stable P2 orbit is born, also along $c_2=0$. 
Figure~\ref{fig:fig1} summarizes all period-$2$ bifurcation loci.
In the small diamond-shaped region in Fig.~\ref{fig:fig1}, 
delimited by  $c_3=0$, $c_4=0$, $c_5=0$ and the Naimark-Sacker line
(indicated by NS),
there is coexistence of  $P2$ and $P2D$ motions. 
On the NS line the $P2$ orbit disappears, to give rise to two closed curves 
(limit cycles).


  \begin{figure}[hbt]
  \begin{center}
  \includegraphics*[width=13cm]{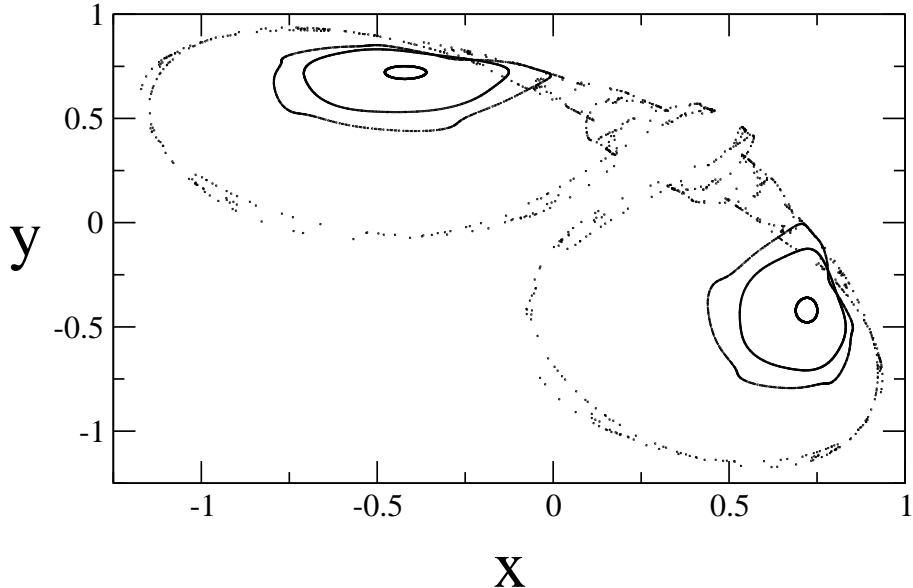}
  \end{center}
\caption{\protect Quasiperiodicity and chaos seen in the
          phase-space of two linearly coupled quadratic maps.
          For $a=0.5$ and $(x_0,y_0)=(0.1234,0.005)$, $1000$ points 
          have been plotted, after a transient of $400$ iterations. 
          Shown are three quasiperiodic solutions, for $b=-0.35$, $-0.38$,
          $-0.40$. 
          Here, smaller $b$ corresponds to larger radius. 
          A frequency-locked solution with period-$6$ appears at 
         $b=-0.41$ (not shown in the figure). 
         The complicated dotted structure is a chaotic orbit for $b=-0.46$.}
  \label{fig:fig3}
  \end{figure}

The stability domains for the coupled quadratic maps can be generated 
numerically, using the method described in Ref.~\cite{g93}, 
and they are shown in Fig.~\ref{fig:fig2}, displaying the richness
of the several coexisting motions in parameter space.
In this figure the basin of unbounded (diverging) attractors is indicated 
by $\infty$. 
The lower border line seen between the P2 domain and black domain is
defined by the Naimark-Sacker line. 
This means that inside the black domain, immediately below the P2 region, 
we have quasiperiodic motion with two limit cycles originated from
P2  trajectories. As we penetrate more and more into this black region, 
we reach a region of chaos. 
The same happens for P4 oscillations,
the difference in this latter case being that the region of quasiperiodic 
motions has four limit cycles originated from P4 orbits.

Figure~\ref{fig:fig3} shows the birth and evolution 
of the Naimark-Sacker bifurcation in phase space
when we walk along the vertical line $a=0.5$ in  parameter space. 
Here we plotted the amplitudes of the maps for four representatives
values of the parameter $b$. 
For $b$ close to the Naimark-Sacker line the motion covers a closed curve 
and is quasiperiodic. For smaller values of $b$, 
a quasiperiodic transition to chaos takes place. 
Periodic windows, not visible in Fig.~\ref{fig:fig3}, appear in this route 
to chaos via quasiperiodicity.


We demonstrated that two linearly coupled quadratic maps 
can show quasiperiodic motion arising from a Naimark-Sacker bifurcation 
of a P2 orbit. 
This was shown analytically and corroborated through a numerical
simulation, as may be seen from the
similarity between Figs.~\ref{fig:fig1} and \ref{fig:fig2}. 
The first figure was obtained analytically while the second numerically.
The precise analytical determination of the Naimark-Sacker road to stable
quasiperiodic motions opens the possibility of several numerical experiments,
in particular, to applications connected with synchronization in networks
of maps\cite{ocean1,ocean2,ocean3}.
Another interesting  application is to investigate how a mismatch in the 
parameters would affect the bifurcation structure.
These applications will be reported elsewhere.

\end{document}